\documentclass[11pt]{article}
\usepackage{graphicx}
\usepackage{amsmath}
\usepackage{amssymb}

\begin{document}



\title{
  The Rock--Paper--Scissors Game\thanks{H.-J. Zhou, ``The rock-paper-scissors game'', Contemporary Physics 57, 151--163 (2016).
    Doi:10.1080/00107514.2015.1026556}
}

\author{Hai-Jun Zhou\\
  {\em{Institute of Theoretical Physics, Chinese Academy of Sciences,}} \\
     {\em{Zhong-Guan-Cun East Road 55, Beijing 100190, China}}
}

\date{\empty}
\maketitle

\begin{abstract}
  Rock-Paper-Scissors (RPS), a game of cyclic dominance, is not merely a popular children's game but also a basic model system for studying decision-making in non-cooperative strategic interactions. \\
  Aimed at students of physics with no background in game theory, this paper introduces the concepts of Nash equilibrium and evolutionarily stable strategy, and reviews some recent theoretical and empirical efforts on the non-equilibrium properties of the iterated RPS, including collective cycling, conditional response patterns, and microscopic mechanisms that facilitate cooperation. We also introduce several dynamical processes to illustrate the applications of RPS as a simplified model of species competition in ecological systems and price cycling in economic markets. 
\end{abstract}

\section{Introduction}

Statistical mechanics aims at understanding collective behaviours of
many-particle systems from microscopic interactions \cite{Huang-1987}. If the
system is a  physical one, each interaction among a subset of particles is
associated with an energy, and the total energy of the system is simply the
sum of all these energies. The system prefers to stay in microscopic
configurations that minimize the total energy, but it is constantly disturbed
by the environment. This competition between energy minimization and
environmental perturbation leads to very rich non-equilibrium dynamics and to
many equilibrium phase transitions in the system's macroscopic property
\cite{Huang-1987}.

Various collective behaviours, driven by strategic interactions among selfish
agents, also emerge in game systems. As a new research field of statistical
mechanics, exploring and understanding the complex non-equilibrium properties
of such competitive social systems became rather active in recent years 
\cite{Szabo-Fath-2007,Castellano-Fortunato-Loreto-2009,Roca-Cuesta-Sanchez-2009,Frey-2010,Perc-etal-2013,Szolnoki-etal-2014,Chakraborti-etal-2015,Huang-2015}.
Yet statistical mechanical approaches to strategic interactions,
compared with physical systems, are facing two additional major challenges.

In a game system, each interaction brings a payoff to every involved agent,
but a fundamental distinction with physical systems is that the payoffs for
different agents of the same interaction  are in general different. Such
differential payoffs cause all the conflicts and competitions in the system
\cite{Osborne-Rubinstein-1994}. Every agent seeks to maximize its own payoff,
but the increase of one agent's payoff does not necessarily mean an increase of
the sum of payoffs of all the agents. Since the  microscopic dynamics is not
guided by the total payoff, the conventional concept of equilibrium Boltzmann
distribution of the total payoff is not useful.

Another major challenge is that the microscopic mechanisms of decision-making
are quite unclear. In many game systems the agents have certain degree of
intelligence, and they make decisions in complicated ways taking into account
both the past experiences and the anticipated future events. Furthermore, the
microscopic parameters of decision-making may evolve in time as a result of
learning and adaptation. 

Various games have been investigated in the literature, of which
the most widely discussed probably is the Prisoner's Dilemma game, devised
by Albert Tucker about 60 years ago \cite{Rapoport-Chammah-1965,Fisher-2008}.
This game is a paradigm for studying cooperation of selfish agents and is the
focus of thousands of research papers \cite{Axelrod-1984,Hauert-Szabo-2005}. On
the other hand the Rock-Paper-Scissors (RPS) game is a paradigm for studying
competition caused by cyclic dominance \cite{Szolnoki-etal-2014}, yet it is much
less discussed. In the present paper we review some aspects of the RPS game 
for physics students, assuming the reader has no background in
game theory.

\begin{figure}[t]
  \centering
    \includegraphics[width=0.5\textwidth]{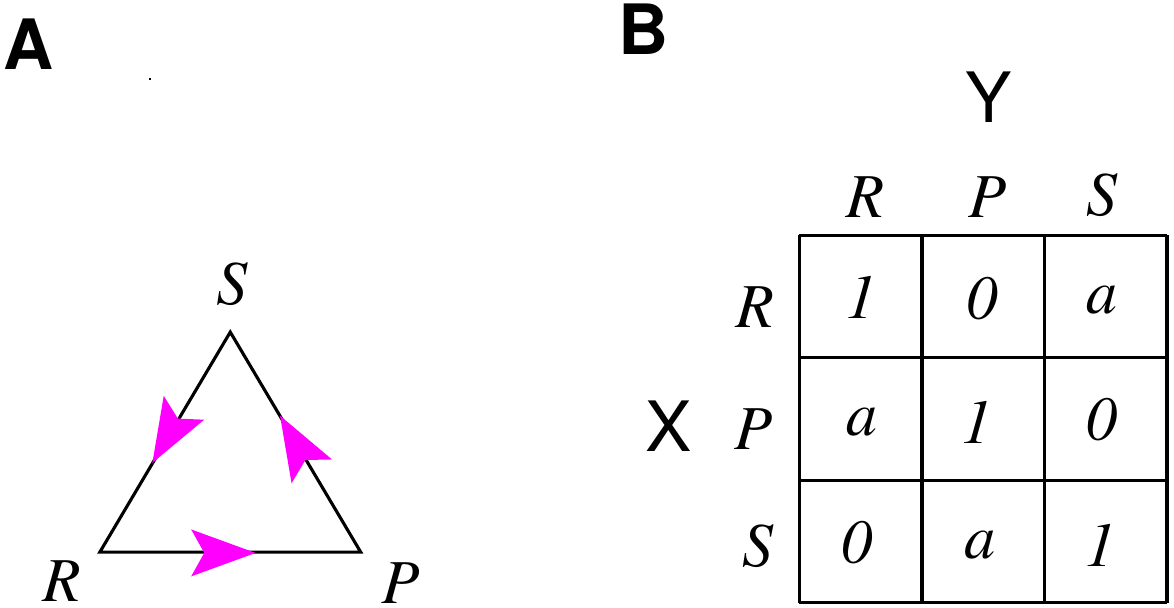}
   \caption{\label{fig:RPSmodel}
    The Rock-Paper-Scissors game. (A) cyclic dominance among the three actions
    $R$, $P$ and $S$, where an arc $s_1 \rightarrow s_2$ from action $s_1$ to
    action $s_2$ indicates that $s_2$ beats $s_1$. (B) The payoff matrix. Each
    matrix element is the payoff of the row player $X$ in action
    $s \in \{R, P, S\}$ against the column player $Y$ in action
    $s^\prime \in \{R, P, S\}$. This figure is adapted from \cite{Bi-Zhou-2014}.
  }
\end{figure}

Being a popular game, the origin of RPS has been difficult to
trace, but there is some written evidence suggesting the Chinese played it
already in the Han Dynasty more than two thousand years ago. It is the simplest
competition system manifesting the ancient cyclic-dominance concept of oriental
taoism philosophy. This game brings fun to people of all ages and occasionally
serves as a fair mechanism to resolve choice conflicts among friends or family
members. There are three possible action choices: $R$ (rock), $P$ (paper) and
$S$ (scissors). Action $R$ is better than $S$, which in turn is better than $P$,
which in turn is better than $R$ (Figure~\ref{fig:RPSmodel}A). Because of this
cyclic dominance none of the three actions is an absolute winning choice. In the
simplest case  the game is played by two players and they compete
simultaneously. For example if one player X chooses $R$ while the other player
Y chooses $S$, then X is the winner.

The RPS game is not merely a game for fun, it actually has fundamental
importance as a basic model system for non-cooperative strategic interactions.
In theoretical studies, cyclic dominance is expressed more quantitatively by a
payoff matrix. A frequently used one is shown in Figure~\ref{fig:RPSmodel}B with
the parameter $a$ being the reward of the winning action. For example, if player
X chooses action $R$ and her opponent Y chooses $S$ then the payoff to X is $a$
while that to Y is $0$; if both players choose the same action (e.g., $P$ versus
$P$) then a tie occurs and each player gets unit payoff. To ensure the property
of cyclic dominance we require $a > 1$. At the specific value $a=2$ the total
payoff of the two players is the same no matter whether the output is win-lose
or tie. If $a>2$, win-lose offers a higher total payoff than tie, while the
reverse is true for $1<a<2$. In more general payoff matrices the winning payoffs
of the three different actions are different, then the rotational symmetry among
$R$, $P$, and $S$ are broken (an example will be given in
Section~\ref{sec:price}).

In the following sections, we first introduce the concepts of Nash equilibrium
and evolutionarily stable strategy for the RPS game (Section~\ref{sec:Nash}), 
and then discuss some recent theoretical and empirical efforts on the
non-equilibrium properties of iterated RPS, such as collective cycling,
conditional response patterns, and microscopic mechanisms that facilitate
cooperation (Sections~\ref{sec:Nash} and \ref{sec:human}). The applications of
RPS in understanding species competition of ecological systems and price cycling
of economic markets are discussed  in Sections~\ref{sec:species} and
\ref{sec:price}. We conclude this review with a brief outlook in 
Section~\ref{sec:outlook}.

\section{Nash Equilibrium, Evolutionary Stability, and Cooperation-Trap Strategies}
\label{sec:Nash}

There are two general theoretical frameworks to study strategic interactions,
classic game theory \cite{Nash-1950,Osborne-Rubinstein-1994} and evolutionary
game theory \cite{MaynardSmith-Price-1973,MaynardSmith-1982,Weibull-1995}.
Classic game theory is based on the assumption that the players
can make completely rational decisions, while evolutionary game theory tries
to understand game outcomes from the angle of evolution and adaptation.
Consider a simple scenario of two players X and Y repeating RPS for an
indefinite number of rounds. The payoff matrix is identical to that of
Figure~\ref{fig:RPSmodel}B, with $a$ being a constant. What shall we expect to
observe concerning this system's long-time behaviour? Classic game theory
gives a clear-cut answer that it will reach a unique mixed-strategy Nash
equilibrium,  while evolutionary game theory is more cautious, emphasizing
that the actual microscopic dynamics of decision-making is also a crucial
factor.

\subsection{Nash-equilibrium mixed strategy}

Classic game theory assumes that the players all have unbounded rationality,
and the solution concept of Nash equilibrium (NE) plays a
fundamental role in this theoretical framework. In an iterated RPS between
two such players, if player X always chooses the same action (say $R$), the
other player Y naturally will always choose the winning action ($P$) to maximize
his own payoff. If player X always chooses between two actions (say $R$ and
$S$), player Y will also take advantage of such a regularity and beat X by
always choosing between actions $R$ and $P$. To avoid being exploited, player X
should therefore adopt all the actions with positive probabilities. But how?

Since any regularity of choices might be detected and be exploited by the
rational (and intelligent) player Y, it is safe for player X to make an action
choice in each game round completely independent of her choices in the previous
rounds. Let us denote by $w_r$, $w_p$ and $w_s$ the respective probability of
$R$, $P$ and $S$ being chosen by player X in one game round
($w_r + w_p + w_s \equiv 1$). The probability vector $(w_r, w_p, w_s)$ is
referred to as a mixed strategy in the game theoretical literature
\cite{Osborne-Rubinstein-1994}. (If two of the action probabilities are
strictly zero, a mixed strategy reduces to a pure strategy.) Under a given mixed
strategy of X, if player Y takes action $R$ then his expected payoff $g_r$ per
game round is simply $g_r = w_r + a w_s$. Similarly the expected payoffs $g_p$
and $g_s$ of $P$ and $S$ are $g_p = w_p + a w_r$ and $g_s = w_s + a w_p$.

Notice that if player X chooses a particular mixed strategy $(w_r, w_p, w_s)$
with
\begin{equation}
  \label{eq:mix13}
  w_r = w_p = w_s = \frac{1}{3} \; ,
\end{equation}
the expected payoff of player Y is independent of his action and
$g_r = g_p = g_s = g^0$ with
\begin{equation}
  \label{eq:NErps}
  g^0 = \frac{1+a}{3} \; .
\end{equation}
In other words, it is impossible for Y to exploit X if the
  latter completely randomize her action choices. Similarly if player Y sets his
mixed strategy to be $(1/3, 1/3, 1/3)$, then the expected payoff of player X is
equal to $g^0$ and it can not be further increased no matter how hard X tries to
adjust her mixed strategy. That is, the chance of player X to take advantage of
Y is also eliminated.

Equation (\ref{eq:mix13}) is a NE mixed strategy for the two-person iterated
RPS. In general, for a game with two or more action choices and involving two or
more players, we define an action probability vector of a player $i$ as a NE
mixed strategy for this player if $i$ is unable to increase its expected payoff
by changing to any another mixed strategy when all the other players keep their
own mixed strategy. If every player of the population is taking such a mixed
strategy the whole system is then said to be in a Nash equilibrium
\cite{Osborne-Rubinstein-1994}.

It can be easily checked that the two-person iterated RPS has only a unique NE
at any $a>1$ and, the probability vector $(1/3, 1/3, 1/3)$ is the only NE mixed
strategy. We can extend the discussion to the scenario of more than two players.
At each game round every player competes with all the other players or only with
a single randomly chosen player. If we consider only probability vectors
$(w_r, w_p, w_s)$ that are strictly mixed (satisfying $w_r w_p w_s > 0$), then
it is relatively easy to prove that $(1/3, 1/3, 1/3)$ is also the unique NE
strictly mixed strategy \cite{Wang-Xu-Zhou-2014}. 

The strategy $(1/3, 1/3, 1/3)$ is maximally random. When a system of $N$ players
reaches the Nash equilibrium, all the $3^N$ possible microscopic configurations
are equally likely to be observed, and the entropy of the system achieves the
global maximum value. Since every player makes decisions independent of other
players and of the previous decisions, the dynamical property of the system is
completely trivial.

The mixed-strategy Nash equilibrium, although being unique for the iterated RPS
game, may not necessarily be stable under small perturbations. Since different
action choices bring the same expected payoff $g^0$ to a player $i$, the
strategy of this player may drift away from  (\ref{eq:mix13}), 
which then will trigger strategy adjustments from the other players. Facing
these induced deviations, if it is the best response for player $i$ to deviate
further away from (\ref{eq:mix13}) then the Nash equilibrium is unstable. To
investigate this type of local stability one often needs to specify how
individual players update their action choices. A comprehensive review on
population dynamics of strategic interactions is presented in
\cite{Sandholm-2010}.
 
In the next subsection we introduce another type of stability criterion which
is independent of the particular microscopic competition dynamics.

\subsection{Evolutionary stability}
\label{sec:ess}

Stability of a mixed strategy can also be defined under the perspective
of mutation and selection
\cite{MaynardSmith-Price-1973,MaynardSmith-1982,Weibull-1995}.
Let us consider a population of $N$ agents interacting with a mixed strategy,
say $\vec{w}$. Suppose now a mutation occurs to a subpopulation of $n < N$
agents such that these $n$ agents adopt a different strategy (say 
$\vec{w^\prime}$). For this hybrid system, if the expected payoff $g$ of an
agent in the unperturbed subpopulation is higher than the expected payoff
$g^\prime$ of an agent in the mutated subpopulation, then the original strategy
$\vec{w}$ is regarded as an evolutionarily stable strategy, otherwise it is an
evolutionarily unstable strategy \cite{Taylor-Jonker-1978}. The concept of
evolutionary stability is basic to evolutionary game theory and
is very useful for understanding biological evolution
\cite{MaynardSmith-1982,Dawkins-2006}.

\begin{figure}[t]
  \centering
  \includegraphics[width=0.5\textwidth]{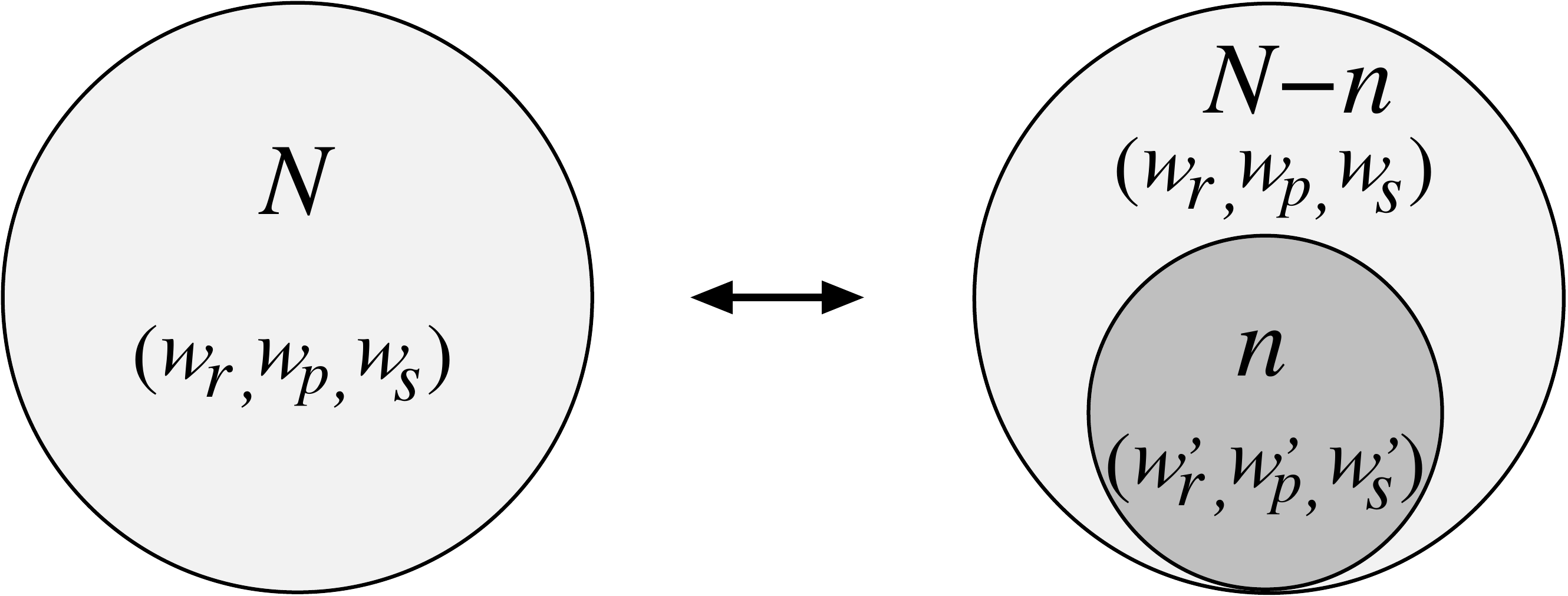}
  \caption{\label{fig:EES}
    Evolutionarily stable strategy for the RPS game. (left) A population of $N$
    individuals adopting a mixed strategy $(w_r, w_p, w_s)$. (right) A mutation
    occurs to $n$ individuals, which form a subpopulation adopting a mutated
    strategy $(w_r^\prime, w_p^\prime, w_s^\prime)$. If the expected payoff of an
    individual in the non-mutated subpopulation is higher than that of an
    individual in the mutated subpopulation for any mutated strategy, the mixed
    strategy $(w_r, w_p, w_s)$ is an evolutionarily stable strategy, otherwise
    it is evolutionarily unstable.
  }
\end{figure}

For the iterated RPS game, let us assume the mutated strategy of $n$ members is
$(w_r^\prime, w_p^\prime, w_s^\prime)$ while the remaining $(N-n)$ members adopt
the NE mixed strategy $(w_r, w_p, w_s) = (1/3, 1/3, 1/3)$, see
Figure~\ref{fig:EES}. At each game round every player competes with a randomly
chosen member of the whole population. The expected payoff for a player in the
unperturbed subpopulation is simply $g = g^0$, while that of a player in the
mutated subpopulation is
\begin{equation}
  g^\prime = g^0 - \frac{(a-2) n}{2 N}  \Bigl[ (w_r^\prime - 1/3)^2 
    + (w_p^\prime - 1/3)^2 + (w_s^\prime - 1/3)^2 \Bigr] \; .
\end{equation}

If $a > 2$ we have $g> g^\prime$, players adopting the NE strategy have a higher
expected payoff than players adopting the mutated strategy. By natural selection
the mutated subpopulation should shrink in size ($n\rightarrow 0$), making the
NE strategy evolutionarily stable. On the other hand, if $1<a<2$ we have
$g < g^\prime$, then the NE strategy is not evolutionarily stable and can not
persist under strategy mutations. In this latter parameter region the system
actually has no evolutionarily stable strategy. 

Why is the strategy $(1/3, 1/3, 1/3)$ evolutionarily stable when
$a>2$ but unstable when $a<2$? The reason lies in the interactions within the
mutated subpopulation. Consider two players adopting the mutated strategy. If
they meet, the probability of tie is 
$$
\frac{1}{3} + \Bigl[ (w_r^\prime-1/3)^2 + (w_p^\prime - 1/3)^2 +
(w_s^\prime  - 1/3)^2 \Bigr] \; ,
$$
which is larger than $1/3$. As the mean payoff $a/2$ of a win-lose output is
less than that of a tie output only for $a<2$, the mutated strategy is
beneficial only in this parameter region.

In the above discussions, a mixed strategy $(w_r, w_p, w_s)$ is defined at the
level of individual players. We can also define a mixed strategy at the
population level. In this latter perspective each individual may hold a fixed
action (a pure strategy), then the mixed strategy describes the fractions of
individuals adopting the different actions, which is more appropriate for
studying strategic interactions in some biological systems. The same analysis
of evolutionary stability can be carried out  at the population level to answer
the question of stable population composition under natural selection
\cite{MaynardSmith-1982,Dawkins-2006}.

\subsection{Cooperation-trap strategies}
\label{sec:ctstrategy}

If an iterated RPS system is in the mixed-strategy Nash equilibrium, every
player stays in a safe position free of being exploited by the others. This is
of course fine in terms of risk avoidance, but on the other hand as all players
are interacting non-cooperatively they might miss the opportunity of achieving
higher accumulated payoffs. Is it possible to sustain high degree of cooperation
in this intrinsically non-cooperative game and beat the Nash equilibrium? For
the two-person iterated RPS, it was demonstrated in \cite{Bi-Zhou-2014} that
there do exist simple strategies that are maximally fair and also maximally
profitable to both players. Such strategies are referred to as cooperation-trap
strategies \cite{Bi-Zhou-2014} as they can induce an opponent player into
complete cooperation. Here we offer an implementation of cooperation-trap
strategies that improves the original protocol suggested in \cite{Bi-Zhou-2014}.

\begin{figure}[t]
  \centering
    \includegraphics[angle=270,width=1.0\textwidth]{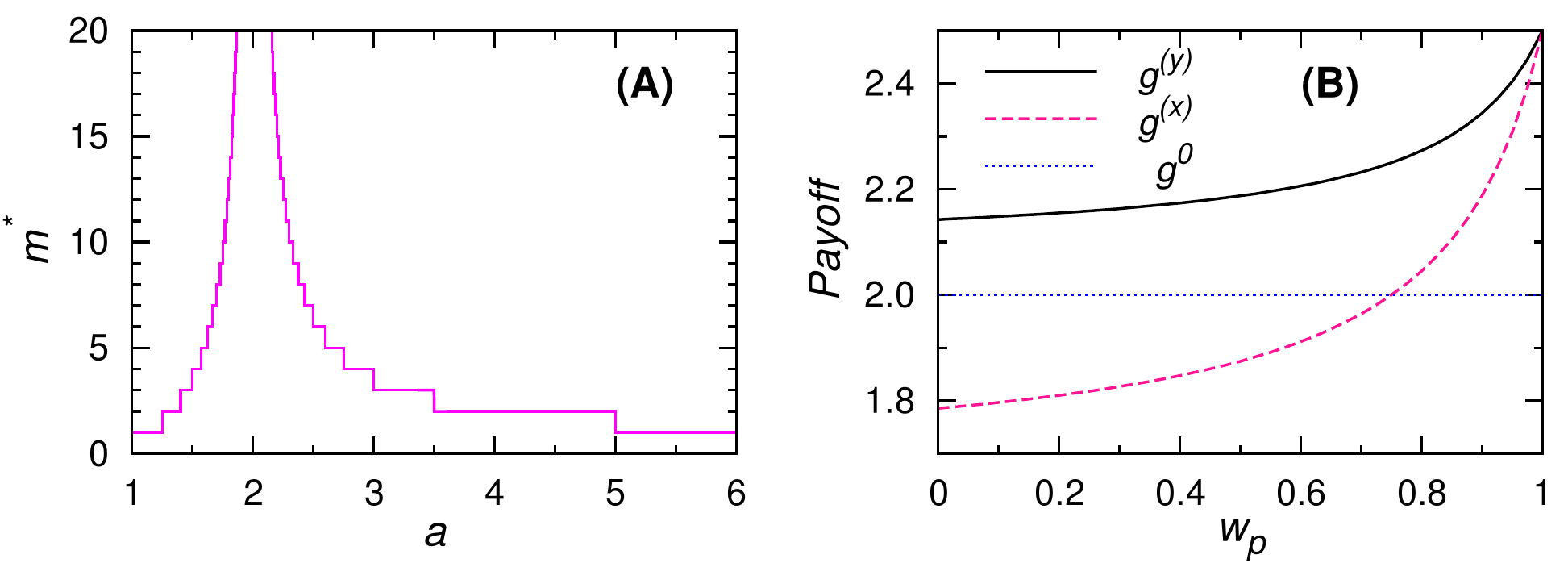}
  \caption{\label{fig:CTrps}
    The cooperation-trap strategy. (A) The minimal memory length $m^*$ as a
    function of the payoff parameter $a$. $m^* \propto |a-2|^{-1}$ in the
    vicinity of $a=2$, and $m^*\equiv 1$ for $a>5$. (B) If player X employs the
    cooperation-trap strategy of memory length $m=6$ while player Y employs a
    mixed strategy $(1-w_p, w_p, 0)$, the expected payoffs per game
    round for X ($g^{(x)}$) and Y ($g^{(y)}$) as compared to $g^0$ of the
    Nash-equilibrium mixed strategy ($a=5.0$). }
\end{figure}

When the reward parameter $a>2$ the mean payoff to a player from a win-lose is
higher than the payoff of a tie.  An intelligent and rational player (say X)
therefore has incentive to search for a strategy that maximizes the chance of
wins while minimizes the chance of ties. The recipe of a simple cooperation-trap
strategy goes as follow:
\begin{enumerate}
\item By default, player X acts in the \emph{cooperation mode}. In this default
  mode, X avoids using one of the three actions (say $P$) and adopts the
  remaining two actions ($R$ and $S$) with equal probability $1/2$ in each game
  round. If player Y cooperatively responds to X by adopting action $P$, then
  both players get an equal expected payoff $a/2$ per round, which is a fair
  result and is higher than the value $g^0$ of the NE mixed strategy.
\item However if player Y exploits the cooperation mode of X by adopting action
  $R$, which returns an even higher expected payoff $(1+a)/2$ to Y, player X
  switches to the \emph{punish mode} in the next $m$ game rounds. In this punish
  mode X employs the NE mixed strategy and adopts actions $R$, $P$ and $S$ with
  equal probability $1/3$. The expected payoff per round is then reduced to
  $g^0$ for both players. Player X is forgivable and she switches back to the
  default cooperation mode after each punish mode of length $m$.
\end{enumerate}
If player X employs this cooperation-trap strategy, it is beneficial for player
Y to abandon action $S$; furthermore if the punish mode duration $m$ is equal to
or larger than a minimum value $m^* \equiv \lceil 3/(a-2) \rceil$, then it is
optimal for Y to fix his action to $P$ in every game round. When
$a>5$ we have $m^*=1$, suggesting that a minimum punish level is enough
to sustain complete cooperation (Figure~\ref{fig:CTrps}A).

When $1<a<2$ the mean payoff of a win-lose is less than the tie payoff. In this
region the default cooperation mode of player X is modified to promote tie
rather than win-lose. By default player X adopts the same action (say $P$) in
every game round as long as player Y responds also with action $P$. If Y
exploits X by adopting action $S$ in one game round,  X switches to the punish
mode in the next $m$ rounds and then switches back to the cooperation mode. It
is also easy to verify that if $m \geq m^* \equiv \lceil 3(a-1)/(2-a)\rceil$
complete cooperation between the two players can be achieved
(Figure~\ref{fig:CTrps}A).

To achieve high degree of cooperation we need a proactive player X to initialize
cooperation. Does the opponent player Y also need to be sufficiently intelligent
to figure out the intention of X? This may not be necessary. Let us consider a
totally myopic player Y who employs a mixed strategy $(w_r, w_p, w_s)$ and
changes this strategy in time to maximize his gain. Since action $R$ is better
than $S$ when facing a player X adopting the cooperation-trap strategy, $w_s$
will evolve to zero and then the mixed strategy of Y becomes $(1-w_p, w_p, 0)$. 
Player Y's expected payoff $g^{(y)}$ per game round is compared with the
corresponding $g^{(x)}$ of player X and the NE payoff $g^0$ in
Figure~\ref{fig:CTrps}B for the case of $a=5$ and $m=6$. We notice that
$g^{(y)}$ is a strictly increasing function of $w_p$, indicating that complete
cooperation ($w_p=1$) is the only fixed point of any gradual learning process
of player Y.

Figure~\ref{fig:CTrps}B also demonstrates that the expected payoff of player X
is less than that of Y unless complete cooperation is reached. Indeed $g^{(x)}$
is even less than the NE payoff $g^0$ if player Y is not sufficiently
cooperative. Player X may overcome such a drawback by further refinements of
the cooperation-trap strategy. This issue and empirical evaluations of the
cooperation-trap strategies will be further studied in a systematic way.

\section{Conditional Response Patterns in Human Subjects}
\label{sec:human}

A convenient assumption of the preceding section is that players of infinite
rationality (who must have excellent random-number generators!) make decisions
based on certain mixed strategy $(w_r, w_p, w_s)$ and modify this strategy
according to feedback information from the iterated RPS. However such an
assumption is often not realistic in competition processes involving human
subjects. A person may not be well conscious of a mixed strategy, rather the
decision-making is heuristic and is easily disturbed by environmental and
psychological factors (e.g., automatic imitation of opponent's actions
\cite{Cook-etal-2012}). Even if the human brain may have the capacity of
implementing a mixed strategy, generating a random sequence of actions following
this strategy is itself very demanding (the human brain performs poorly in
randomization tasks \cite{Wagenaar-1972,Jahanshahi-etal-2006}).

How do people choose actions in the iterated RPS? As a first step to answer this
difficult question, empirical investigations were carried out in the last few
years by several research teams
\cite{Hoffman-etal-2012,Xu-Zhou-Wang-2013,Cason-Friedman-Hopkins-2014,Wang-Xu-Zhou-2014}. 
Two of these experiments \cite{Hoffman-etal-2012,Cason-Friedman-Hopkins-2014}
employed the all-to-all protocol with every person playing simultaneously
against all the other players in the population, while
\cite{Xu-Zhou-Wang-2013,Wang-Xu-Zhou-2014} employed the more traditional random
pairwise-matching protocol, with every person only competing with another single
player. Here we focus on the results of the latter as the experimental setting
mimics decision-making under uncertainty.

\begin{figure}[t]
  \centering
    \includegraphics[width=1.0\textwidth]{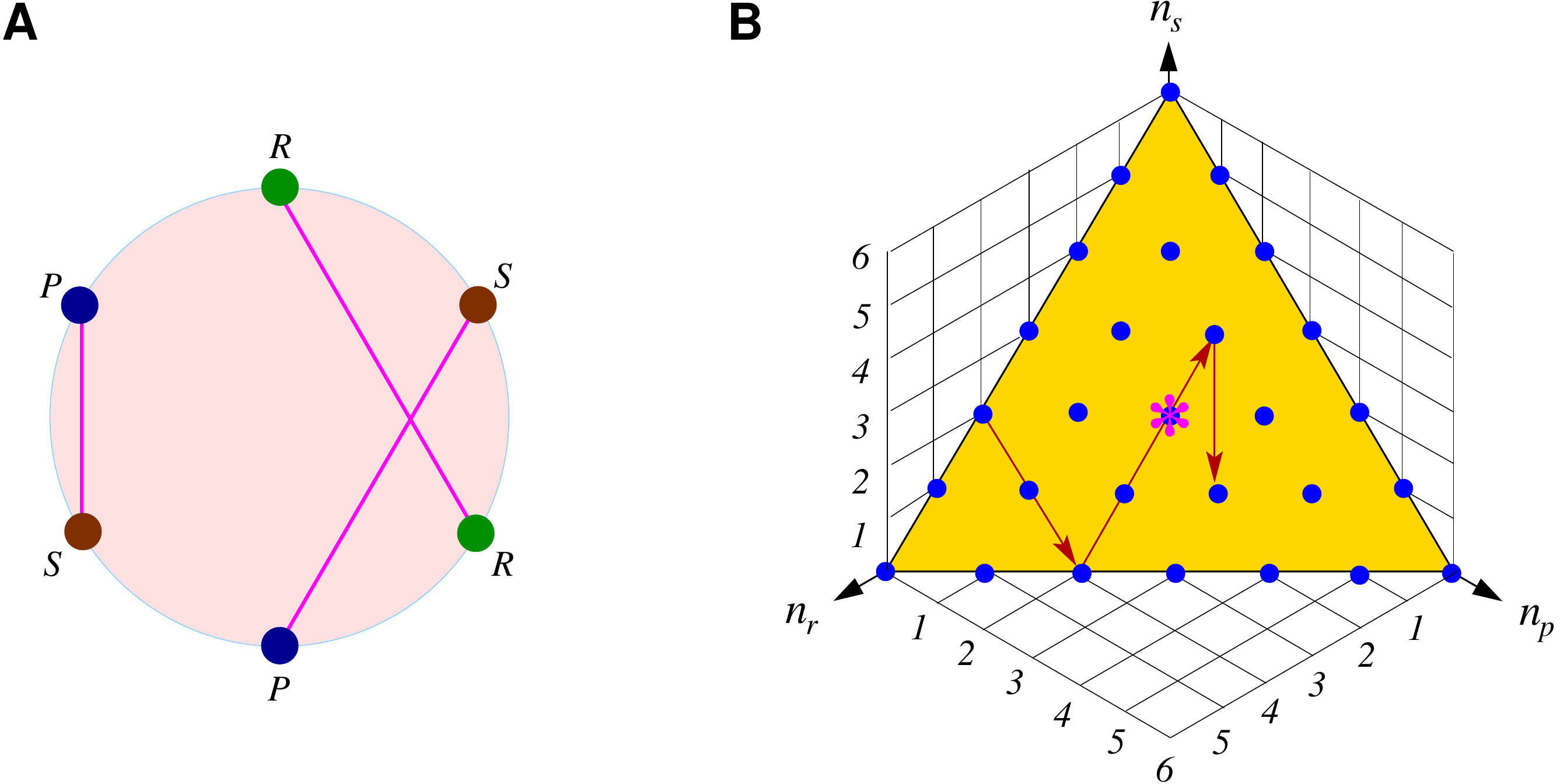}
   \caption{\label{fig:GroupRPS}
    Iterated RPS played by $N=6$ human subjects. (A) The players at each game
    round form $N/2$ random pairs and they play the game once with the pair
    opponent only. The winner of each pair gets payoff $a$ and the loser $0$,
    while the payoff for a tie is $1$. (B) Each social state $(n_r, n_p, n_s)$
    is a point of the triangle confined by $n_r+n_p+n_s=N$ and
    $n_r n_p n_s \geq 0$, where $n_r$ denotes the total number of players
    choosing action $R$ (similarly for $n_p$ and $n_s$). The social state for
    the example of (A) is $(2, 2, 2)$ and is marked by the star symbol at the
    triangle's center. As the game is repeated it leaves a trajectory in the
    social-state triangle. For the shown trajectory segment, the rotation angles
    (with respect to the centroid) of the three social-state transitions are,
    respectively, $\theta=+60^\circ$, $\theta=0^\circ$ and $\theta= - 120^\circ$.
    This figure is adapted from \cite{Wang-Xu-Zhou-2014}. }    
\end{figure}

A total number of $360$ university students participated in the experiment of
Wang and Xu \cite{Xu-Zhou-Wang-2013,Wang-Xu-Zhou-2014}. These human subjects
form $60$ groups with each group (population) containing six people. As shown in
Figure~\ref{fig:GroupRPS}A, at each game round $t=1, 2, \ldots, 300$, the six
players of each population are randomly paired and they play once with their
pair opponent (which might be different in different rounds) under the payoff
rule of Figure~\ref{fig:RPSmodel}B, and then every player receives feedback
information about her/his own payoff in this round, her/his accumulated payoff, 
and the opponent's action in this round.

In the experiment each population plays with a fixed 
reward parameter $a$ whose value ranging from $a=1.1$ to $a=100$.
The empirical results demonstrated that the precise value of $a$ does not
affect the qualitative dynamical behaviour of this finite-population
system \cite{Wang-Xu-Zhou-2014,Wang-Xu-2014}.

\subsection{Individual inertia effect and collective cycling}
\label{subsec:RPSobserv6}

\begin{figure}[t]
  \centering
  \includegraphics[angle=270,width=1.0\textwidth]{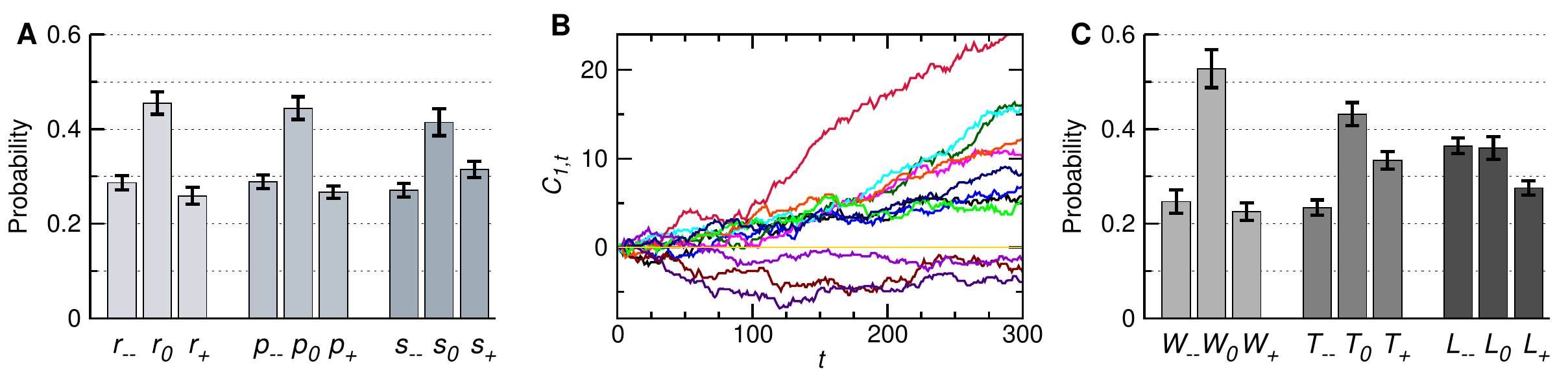}
  \caption{\label{fig:RPSexp}
    Statistical regularity of the iterated RPS with payoff parameter $a=2.0$
    played by $72$ human subjects (divided into twelve groups of size $N=6$).
    (A) The mean action shifting probability of a player (and the standard
    error of this mean) conditional on the current action. Given the current
    action being $R$,  the probability of making a clockwise shift
    ($R\rightarrow S$), of repeating action $R$, of making a counter-clockwise 
    shift ($R\rightarrow P$) are denoted as $r_{-}$, $r_0$ and $r_{+}$,
    respectively. The probabilities $p_{-}, p_{0}, p_{+}$ and
    $s_{-}, s_{0}, s_{+}$ are defined similarly. (B) The accumulated cycle
    numbers $C_{1,t}$ in the first $t$ game rounds as obtained for the twelve
    populations. (C) The mean action shifting probability of a player (and the
    standard error of this mean) conditional on the result of the current play
    being win (W), tie (T) or lose (L). Given the current result being W, the
    probability of making a clockwise action shift, of repeating the same
    action, of making a counter-clockwise shift are denoted as $W_{-}$, $W_0$
    and $W_{+}$, respectively. The probabilities $T_{-}, T_{0}, T_{+}$ and
    $L_{-}, L_{0}, L_{+}$ are defined similarly. This figure is adapted from
    \cite{Wang-Xu-Zhou-2014}. }    
\end{figure}

During the $300$-round iteration, each player leaves an action sequence
$(s_1$, $s_2$, $\ldots$, $s_{300})$ with $s_t \in \{R, P, S\}$ being the action at the
$t$-th round. We then get the preference vector of this player to the three
actions, $(f_r, f_p, f_s)$, where $f_r$ is simply the fraction of rounds the
action being $R$ (similarly for $f_p$ and $f_s$, and $f_r+f_p+f_s\equiv 1$).
We find that, consistent with the Nash equilibrium theory, the vectors 
$(f_r, f_p, f_s)$ of the players are close to the mixed strategy
$(1/3, 1/3, 1/3)$. For example $f_r = 0.36 \pm 0.07$, $f_p = 0.32 \pm 0.07$ and
$f_s = 0.32 \pm 0.06$ at $a=2.0$ (mean and standard deviation, obtained by
averaging over $72$ subjects) \cite{Wang-Xu-Zhou-2014}. This observation is of
course most natural given the rotational symmetry among the three actions.

The Nash equilibrium theory also predicts the choices of a player at two 
consecutive game rounds are completely independent of each other. But this is
not what actually happened in the game. Instead there is considerable degree of
temporal correlation within each action sequence. Especially, as demonstrated in
Figure~\ref{fig:RPSexp}A, a player in each round is more likely to repeat the 
action of the previous round than to shift action either in the
counter-clockwise ($R\rightarrow P$ or $P\rightarrow S$ or $S\rightarrow R$) 
direction or in the clockwise ($R\rightarrow S$ or $P\rightarrow R$ or 
$S\rightarrow P$) direction. This inertia effect is strongest at $a=1.1$, it
weakens slightly with the increase of $a$ and is still very significant even at
$a=100$ \cite{Wang-Xu-Zhou-2014}. On the other hand, if a player does make a
change, this change is symmetric in the sense that the probability of shifting
action in the counter-clockwise direction is almost the same as that of shifting
in the clockwise direction.

Although individual action changes from any action do not have directional
preference, the empirical data reveal persistent counter-clockwise cycling in
the collective behaviour of the population. The population's collective state at
each game round $t$ can be described by the vector $(n_r, n_p, n_s)$ with $n_r$
being the total number of players adopting action $R$ (similarly for $n_p$ and
$n_s$). The evolution of this so-called social state with $t$ then draws a
trajectory in the social-state plane (Figure~\ref{fig:GroupRPS}B), which is
highly stochastic. To detect directional motions that signify deviation from
equilibrium, a rotational angle $\theta$ is assigned to each social-state
transition \cite{Wang-Xu-Zhou-2014}. If the transition is associated with a
counter-clockwise rotation around the centroid of the social-state plane, then
$\theta$ is positive; if it is associated with a clockwise rotation around the
centroid, then $\theta$ is negative; in all the other cases $\theta=0$
(Figure~\ref{fig:GroupRPS}B). The net number of turns $C_{1,t}$ a trajectory
cycles around the centroid of the social-state plane from the first to the
$t$-th game round can then be obtained by adding up the rotation angles. As
illustrated in Figure~\ref{fig:RPSexp}B for $a=2.0$, the accumulated cycle
number $C_{1, t}$ has a linearly increasing trend with $t$, revealing persistent
collective cycling along the $R\rightarrow P \rightarrow S\rightarrow R$
direction. The mean cycling frequency is $\nu \approx +0.03$ (one
counter-clockwise cycle in about $35$ game rounds), and this value does not
change significantly with the payoff parameter $a$ \cite{Wang-Xu-Zhou-2014}.

\subsection{The conditional-response mechanism}

The existence of weak but persistent collective cycling
(Figure~\ref{fig:RPSexp}B) is apparently conflicting with the absence of
directionality in the action shift behaviour of individual players
(Figure~\ref{fig:RPSexp}A). Why a seemingly symmetric dynamics at the level of
individual players results in asymmetric motion at the level of the whole
population?

This apparent contradiction has been resolved in \cite{Wang-Xu-Zhou-2014} by
the key observation that players have different degrees of willingness to make
a change, depending on whether the previous play was a win, a tie, or a lose.
Consider the tendencies $W_0$, $T_0$ and $L_0$ of a person to choose the same
action in two consecutive rounds if the earlier round is a win (W), a tie (T)
or a lose (L), respectively. The empirical data reveal $W_0 > T_0 > L_0$, that
is, the tendency of repeating an action increases with the payoff of the earlier
round (Figure~\ref{fig:RPSexp}C). Such a microscopic pattern is qualitatively
similar to the `win-stay lose-shift' strategy of playing the iterated Prisoner's
Dilemma game, whose effectiveness has been confirmed by theoretical calculations
\cite{Grofman-Pool-1977,Kraines-Kraines-1989} and extensive computer simulations
\cite{Kraines-Kraines-1993,Nowak-Sigmund-1993}. The tendencies ($W_{-}$, $T_{-}$
and $L_{-}$) of shifting action in the clockwise direction and the tendencies
($W_{+}$, $T_{+}$ and $L_{+}$) of shifting in the counter-clockwise direction
also depend considerably on the output of the previous round
\cite{Wang-Xu-Zhou-2014}.

Inspired by these empirical conditional response patterns, a simple model of
decision-making was proposed in \cite{Wang-Xu-Zhou-2014} to understand the
interactions in the iterated RPS. This model assumes the action shifting
probabilities in each game round depend only on the game output in the previous
round. For example if a player wins a play, then in the next round she/he has
probability $W_0$ to stick to the same action, probability $W_{-}$ to shift
action in the clockwise direction, and probability $W_{+}$ to shift action in
the counter-clockwise direction. This conditional-response
microscopic rule is characterized by six independent parameters
$(W_{-}, W_{+}, T_{-}, T_{+}, L_{-}, L_{+})$ which can be empirically fixed and
it completely ignores all the possible higher-order complications in the
decision process. Yet very encouragingly, this simple model quantitatively
reproduce all the major experimental results, including the cycling
frequencies and inertia effects measured in the $60$ different populations
(for more details see \cite{Wang-Xu-Zhou-2014}).

It is really interesting and unexpected that complicated decision-making
processes of human subjects are statistically described by such a simple
conditional-response mechanism. However we should also point out that the
conditional-response model is at present only a phenomenological model. A
missing link of basic significance is how the six independent conditional
response parameters evolve in time as a result of learning. It could be that the
conditional-response mechanism is just the result of certain lower-level
learning dynamics. These questions need to be further studied from the
neurobiology side \cite{Glimcher-etal-2009,Janacsek-Nemeth-2012}.

A counter-intuitive theoretical prediction of the conditional response model is
that social-state cycling does not require microscopic asymmetry
in conditional responses: it persists even if $W_{-}=W_{+}$,
$T_{-}=T_{+}$ and $L_{-}=L_{+}$ provided that $W_{0}$, $T_{0}$ and $L_{0}$ are not
all equal to each other \cite{Wang-Xu-Zhou-2014}. This prediction has been
verified by computer simulations, confirming that social cycling is 
indeed an emergent phenomenon.

\subsection{Discussion on the effect of population size $N$}

Under the random pairwise-matching protocol, each player has  probability
$1/(N-1)$ of encountering the same opponent in two consecutive game rounds.
This probability decreases quickly with population size $N$, and consequently
the action choices of different players are less and less correlated as $N$
increases. Since active decision-making is a costly mental process, as the
action correlation decreases with $N$, the incentive for a player to change
action should be weaker and weaker, especially if the player
wins a previous round. Therefore we expect that the players will be less active
in a larger population and the inertia effect of decision-making will be 
stronger; furthermore the win-stay probability $W_0$ will also increase with
$N$.

The $N=2$ iterated RPS is special. As both players know about the opponent's
action history, it is a complete-information system. To avoid being the loser
in this two-person game, both players should be very active and make their
action choices most difficult to predict. We therefore anticipate that, (i) the
inertia effect of individual players will be most weak, and (ii) the conditional
response probabilities will only weakly deviate from $1/3$ and the players may
not prefer to repeat a winning action.

The effect of population size will be systematically explored by laboratory
experiments. These empirical studies may inspire the construction of refined
learning models for the iterated RPS.

\section{Rock-Paper-Scissors in Species Competition}
\label{sec:species}

Cyclic dominance is a ubiquitous phenomenon in ecological systems. For example
the European honeybees invaded the local honeybees after being
introduced to Japan but they were unprepared for the attacks
from Japanese hornets, while the Japanese honeybees have developed a collective
thermal defense mechanism against the hornets as a result of coevolutionary
adaptation \cite{Ono-etal-1995}. Another more quantitative example is the color
polymorphism of male side-blotched lizards \cite{Sinervo-Lively-1996}. Field
measurements revealed the frequencies of the three types of cyclicly dominant
male lizards oscillate with a period of approximately six years and with a
mutual phase shift of about two years \cite{Sinervo-Lively-1996}.
Ecologists believe cyclic dominance to be a key factor
contributing to ecosystem complexity, and they take the RPS game as a basic
model of species competition and coexistence
\cite{Tainaka-2001,Claussen-Traulsen-2008,Szolnoki-etal-2014}.
In this section we briefly review two simple microscopic ecological processes, 
the collision dynamics and the replicator dynamics.

\subsection{The collision dynamics}

We may consider an ecological system formed by three different species $R$, $P$,
and $S$. The total number of individuals in the system is a fixed integer, $N$. 
Among these individuals  $N \rho_r$ belong to species $R$, $N \rho_p$ to species
$P$, and the remaining $N \rho_s$ to species $S$
($\rho_r + \rho_p + \rho_s \equiv 1$). At each time interval $\delta t =2/N$
there is a competition between a randomly chosen pair of neighbouring
individuals. If this competition is between two individuals of the same species,
then both individuals survive. If it is between an individual of species $R$ and
one of species $P$, then with probability $w_{p\leftarrow r}$ the $R$-individual
is displaced by an offspring of the $P$-individual. The parameter
$w_{p\leftarrow r}$ quantifies the dominance degree of the $P$ species to the
$R$ species. Similarly the other two dominance degrees are denoted as
$w_{r \leftarrow s}$ and $w_{s\leftarrow p}$, respectively. Notice that the
expansion of one species is associated with the shrink of another species 
(constant-sum game).

Such a simple collision model and its various extensions have been investigated
by many authors (see, e.g., reviews
\cite{Tainaka-2001,Frey-2010,Szolnoki-etal-2014}).
In the case of a well-mixed population each individual has equal chance of
encountering any another one, then for $N\rightarrow \infty$ the
species frequencies $\rho_r(t)$, $\rho_p(t)$, $\rho_s(t)$ as functions of
time $t$ are governed by \cite{Itoh-1987,Frean-Abraham-2001}
\begin{subequations}
  \label{eq:RPScollision}
  \begin{align}
    \frac{{\rm d} \rho_r}{ {\rm d} t} & = 
    \rho_r \rho_s w_{r\leftarrow s}
    - \rho_r \rho_p w_{p\leftarrow r} \; ,
      \label{eq:RPScollision:R}\\
      \frac{{\rm d} \rho_p}{ {\rm d} t} & = 
      \rho_p \rho_r w_{p\leftarrow r} - \rho_p \rho_s
      w_{s\leftarrow p} \; ,
      \\
      \frac{{\rm d} \rho_s}{ {\rm d} t} & = 
      \rho_s \rho_p w_{s\leftarrow p} -
      \rho_s \rho_r w_{r\leftarrow s}  \; .
  \end{align}
\end{subequations}
Equation (\ref{eq:RPScollision}), as a simple theory of cyclic dominance among 
three species, can be regarded as an extension of the celibrated Lotka-Volterra
equation on the nonlinear predator-prey interactions between two species
\cite{Kot-2001}. This evolution equation has a fixed-point solution of
\begin{subequations}
  \label{eq:collisionfix}
  \begin{align}
    \rho_r^* &= \frac{w_{s\leftarrow p}}{w_{r\leftarrow s}
      + w_{s\leftarrow p} +
      w_{p\leftarrow r}} \; , \\
    \rho_p^* &= \frac{w_{r\leftarrow s}}{w_{r\leftarrow s} 
      + w_{s\leftarrow p} +
      w_{p\leftarrow r}} \; , \\
    \rho_s^* &= \frac{w_{p\leftarrow r}}{w_{r\leftarrow s} 
    + w_{s\leftarrow p} +
    w_{p\leftarrow r}} \; .
  \end{align}
\end{subequations}
Notice that lowering the dominance degree $w_{r\leftarrow s}$ of species $R$ has
an enhancing effect on its fixed-point frequency $\rho_r^*$. This is because
decreasing $w_{r\leftarrow s}$ has the indirect consequence of suppressing the
growth of species $P$ which preys on species $R$.

\begin{figure}[t]
  \centering
  \includegraphics[angle=270,width=0.8\textwidth]{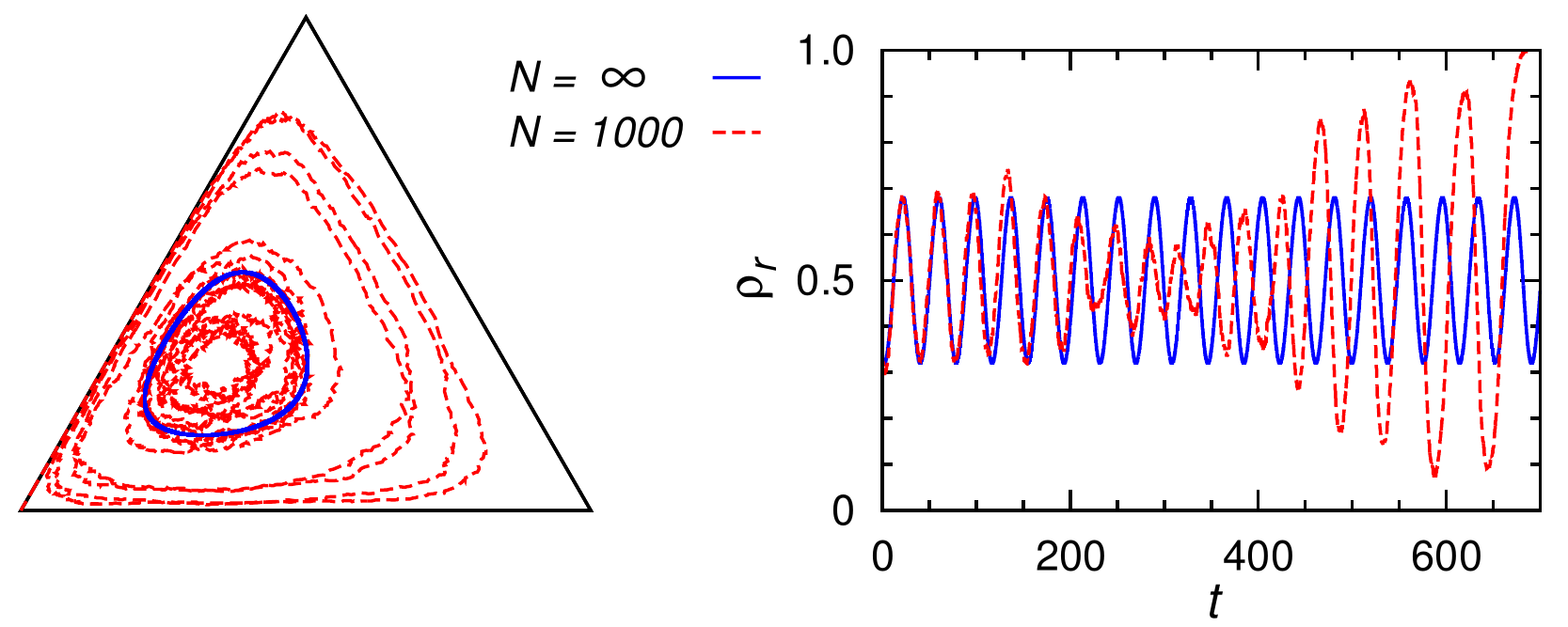}
  \caption{\label{fig:RPScollision}
    The collision model of species competition at dominance parameters 
    $w_{r\leftarrow s}=0.2$, $w_{s\leftarrow p}=0.5$, $w_{p\leftarrow r}=0.3$. Solid
    lines are theoretical results for population size $N=\infty$, dashed lines
    are the results obtained by a single simulation of a population of $N=1000$
    individuals. (left) Trajectory of the population's state
    $\bigl(\rho_r(t), \rho_p(t), \rho_s(t)\bigr)$ starting from the initial
    state $(1/3, 1/3, 1/3)$. The population-state triangle is understood in the
    same way as Figure~\ref{fig:GroupRPS}B. (right) Evolution of $\rho_r(t)$
    with time. }
\end{figure}

If the initial frequencies of the three species deviate from the fixed point
$(\rho_r^*, \rho_p^*, \rho_s^*)$, the system will not evolve into it but will
move around it along a periodic orbit, see Figure~\ref{fig:RPScollision}.
Equation (\ref{eq:RPScollision}) has infinitely many such limiting-cycle
solutions, each of which is characterized by an invariant
$C\equiv (\rho_r / \rho_r^*)^{\rho_r^*} (\rho_p / \rho_p^*)^{\rho_p^*}
(\rho_s / \rho_s^*)^{\rho_s^*}$ \cite{Frean-Abraham-2001,Itoh-1987}.
$C=1$ at the fixed point $(\rho_r^*, \rho_p^*, \rho_s^*)$, and it decreases
continuously to zero as the distance of the limiting cycle to the fixed point
increases.

Since the  population state evolves along a limiting cycle, such an infinite
ecosystem is only marginally stable. This  has a very significant consequence
for a finite ecosystem, namely species extinction is an unavoidable fate.
Figure~\ref{fig:RPScollision} illustrates a single evolutionary trajectory of
a population with $N=1000$ individuals. Starting from an initial state
$\rho_r=\rho_p=\rho_s=1/3$, the population state initially follows the periodic
orbit, it then deviates more and more from this orbit due to the intrinsic
stochasticity of the collision process. At the later stage the oscillatory
magnitudes of the species frequencies become more and more pronounced and
eventually only one species remains in the population. The final surviving
species depends on the whole evolution process, but the species with the lowest
value of dominance degree has the highest chance of survival
(``survival of the weakest'' \cite{Frean-Abraham-2001}).

This prediction of species extinction does not agree with empirical
observations. Different species do coexist in various real-world ecosystems
containing only a finite number of individuals. There have been a lot of
theoretical and experimental studies on this issue. A central aspect of
real-world systems is its spatial structure \cite{Durrett-Levin-1994}. 
Different species cluster into different local regions in a two- or
three-dimensional space, and the interactions between different species occur
only at the boundaries between these regions. Simulation results and
theoretical computations reveal that all the species in such a spatial system
have a high probability of survival even after an infinite evolution time.
The species form very interesting entangled patterns, each region of a species
is a spiral shape, and the boundaries of the regions move in time
\cite{Kerr-etal-2002,Laird-2014}.

In real-world ecosystems, individual animals also move actively in space.
Increased mobility entails increased interactions with the
other species and makes the population be more mixed.
An interesting theoretical observation of \cite{Reichenbach-Mobilia-Frey-2007}
is that there is a sharp `phase transition' of biodiversity in the
Rock-Paper-Scissors ecosystem: if the species mobility exceeds certain critical
value, the probability of species coexistence drops from $\approx 1$ to
$\approx 0$ for a sufficiently large system.
Therefore the RPS ecosystem can tolerate certain degree of species
mobility but not too much.

\subsection{The replicator dynamics}

In many ecological systems, species conflict is not caused by the
direct predator-prey interactions but is the result of 
competing for the same resources.
For example, different strains of budding yeast cells may grow and
reproduce in the same environment. One mother cell divides
into two daughter cells after its body size exceeds
certain critical value, so the number of yeast cells proliferates in an
exponential manner under good nutrient conditions. If one strain of
yeast cells has a higher growth rate than another strain,
it will have a higher reproduction rate and
its population size will then increase exponentially faster than that of
the rival strain \cite{Kerr-etal-2002,Waite-Shou-2012}. 

The interactions among three types of yeast cells were studied in the
experimental system of \cite{Kerr-etal-2002}, and cyclic dominance in
growth-rate advantage was observed. Similar cyclic dominance phenomena 
also exist in other engineered or naturally-occurring microbial ecosystems.
For such an ecosystem, the reproduction rate of a particular species then
depends strongly on the relative abundances of all the species. As a simple
model we may again consider an ecosystem formed by three species $R$, $P$,
and $S$. Let us assume that the reproduction rate $g_r$, $g_p$ and $g_s$
of the three species depend linearly on the species frequencies
$\rho_r$, $\rho_p$ and $\rho_s$:
\begin{subequations}
  \label{eq:rep-rate}
  \begin{align}
    g_r & = c_0 - \rho_p + (a-1) \rho_s \; , \\
    g_p & = c_0 - \rho_s + (a-1) \rho_r \; , \\
    g_s & = c_0 - \rho_r + (a-1) \rho_p \; ,
  \end{align}
\end{subequations}
where the parameter $c_0$ denotes the null reproduction rate when species
competition is absent (for simplicity we assume it is the same for all the
three species), which in general depends on the total number of individuals
in the population. Notice that increasing the frequency $\rho_p$ of the $P$
species has a negative effect on the reproduction rate of the $R$ species,
while increasing $\rho_s$ of the $S$ species has a positive effect on $g_r$
when $a>1$.

\begin{figure}[t]
  \centering
  \includegraphics[angle=270,width=0.6\textwidth]{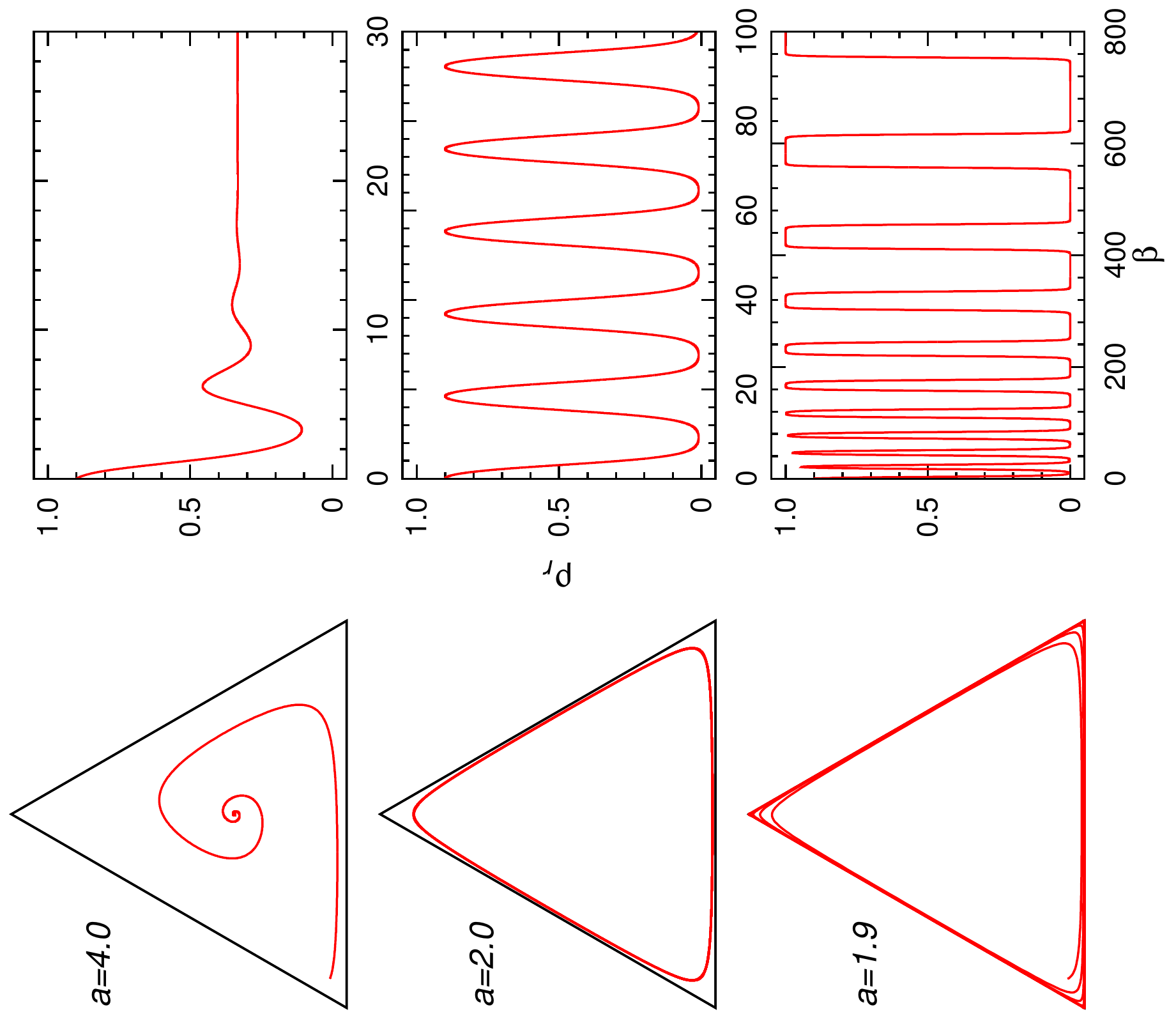}
  \caption{\label{fig:replicator}
    The replicator model of species competition. When $a>2$ (top), the
    population state $(\rho_r, \rho_p, \rho_s)$ evolves towards the steady state
    $(1/3, 1/3, 1/3)$. When $a=2$ (middle), the population state moves along a
    periodic orbit. When $a<2$ (bottom), the population state mores towards the
    boundary of the state space and cycles in a non-periodical manner.
    The left panel shows three representative population state
    evolution trajectories (for $a=4.0$, $2.0$ and $1.9$, respectively)
    starting from the initial population state
    $\rho_r=0.9, \rho_p=\rho_s= 0.05$, while the right panel shows the
    fraction $\rho_r(t)$ as a function of time $t$. }
\end{figure}

The expected number $n_r(t)$ of $R$-individuals evolves with time $t$ according
to ${\rm d} n_r(t) / {\rm d} t = n_r(t) g_r$ and similarly for the other two
values $n_p(t)$ and $n_s(t)$. The total population size
$N(t) \equiv n_r(t) + n_p(t) + n_s(t)$ then evolves according to
${\rm d}N(t) / {\rm d} t = N(t) \overline{g}$, with the mean reproduction rate
being $\overline{g} = \rho_r g_r + \rho_p g_p + \rho_s g_s$. If
$\overline{g}(t)$ is always positive the total population will diverge with
time. Since $\rho_r(t) \equiv n_r(t)/N(t)$ and similarly for $\rho_p(t)$ and 
$\rho_s(t)$, we have
\begin{subequations}
  \label{eq:replicator}
  \begin{align}
    \frac{{\rm d} \rho_r}{{\rm d} t} & =  
    \rho_r \bigl[g_r - \overline{g}\bigr] \; , \\
    \frac{{\rm d} \rho_p}{{\rm d} t} & =  
    \rho_p \bigl[g_p - \overline{g}\bigr] \; , \\
    \frac{{\rm d} \rho_s}{{\rm d} t} & =  
    \rho_s \bigl[g_s - \overline{g}\bigr]  \; .
  \end{align}
\end{subequations}
Equation (\ref{eq:replicator}) is independent of the null reproduction rate
$c_0$. Such an evolution dynamics is usually referred to as the replicator
dynamics \cite{Taylor-Jonker-1978,Weibull-1995}.

The deterministic dynamics (\ref{eq:replicator}) is easy to solve numerically.
This equation has a fixed-point solution of $\rho_r=\rho_p=\rho_s=1/3$. 
If $a>2$, the population state $(\rho_r, \rho_p, \rho_s)$ evolves towards the
fixed point $(1/3, 1/3, 1/3)$ starting from any initial condition satisfying
$\rho_r \rho_p \rho_s > 0$, therefore this fixed point is a globally stable
state and coexistence of all the three species is stable towards perturbations
(Figure \ref{fig:replicator}, top panel).
At $a=2$, however, the evolution converges to a limiting cycle (middle panel
of Figure \ref{fig:replicator}). Similar to the collision model of the preceding
subsection, the ecosystem is then only marginally stable and will eventually
goes to species extinction. To remain biodiversity we need to consider again the
spatial structure of species competition.

When $1< a < 2$, the evolution does not  converge to a fixed point, nor to a
limiting cycle, but keep cycling with longer and longer period. The last
phenomenon of non-periodic oscillation was first discussed in
\cite{May-Leonard-1975}. As shown in the bottom right panel of 
Figure~\ref{fig:replicator}, the frequency of each species (say $\rho_r$) jumps
back and forth between the nearly extinct state ($\rho_r \approx 0$) and the
overwhelmingly occupied state ($\rho_r \approx 1$). The jumps between these two situations occur
very quickly while the residing time in each state become longer and longer.
The reason that a species can recover from nearly
extinction to predominance is mainly due to the assumption of offspring's
exponential proliferation. In the actual situation of a finite population, two
of the species will be extinct inevitably.

The replicator dynamics has been extensively used in game theoretical studies
\cite{Roca-Cuesta-Sanchez-2009}. It is a quantitative model system to discuss 
evolutionarily stable strategies
\cite{MaynardSmith-Price-1973,MaynardSmith-1982}.
This dynamics can also be interpreted as a simple model of social learning
through imitation \cite{Weibull-1995}.

\section{Rock-Paper-Scissors in Market Price Competition}
\label{sec:price}

It is common to observe that different shops in a market sell the same item at
different prices, and furthermore, the price of each shop is not a constant but
changes non-monotonically in time (see, e.g., \cite{Lach-2002,Noel-2007}).
Edgeworth first predicted the existence of persistent price cycles
\cite{Edgeworth-1925}, arguing that under strong competition shops will lower
their prices in small steps to attract more customers, but if the price hits a
bottom level they will suddenly lift price to a much higher level. The RPS game
has been used in the theoretical economics field to qualitatively describe
price dynamics \cite{Hopkins-Seymour-2002}.

Suppose a simplified situation that each day there are $N$ new lazy buyers and
$N$ new diligent buyers looking for a particular item. A lazy buyer enters into
the first shop he found and buys the item, while a diligent buyer examines all
the shops in the market and buys the item at the shop offering the lowest price.
Let us further assume that there are only two shops selling this item and these
two shops can choose to sell at three different price levels, the high price
$h$, the medium price $m$, and the low price $l$.

Initially both shops may sell the item at the high price $h$, and the expected
profit or payoff for each shop is $N h$. If one shop (say X) now lowers the
price to $m$, its expected payoff changes to $\frac{3 N}{2} m$, which is higher
than $N h$ if $m> \frac{2}{3}h$, but the expected payoff for the other shop (Y)
is reduced to $\frac{N}{2} h$. If $l>\frac{2}{3} m$, the best response of shop
Y is to shift price from $h$ to the low value $l$, which will lead to an
expected payoff of $\frac{3 N}{2} l$ for itself and a reduced value
$\frac{N}{2} m$ for shop X. But if $h> 2 l$, shop X will again respond by
shifting to the high price $h$ which increases its expected payoff to
$\frac{N}{2} h$,  ..., causing persistent price oscillations
\cite{Cason-Friedman-Wagener-2005}. The above analysis demonstrates that in the
parameter range of $\frac{2}{3}h <m<\frac{3}{2}l$ and $2 l < h< \frac{9}{4} l$,
there is the cyclic-dominance of the medium price $m$ beating the high price
$h$, the low price $l$ beating the medium price $m$, and the high price $h$
beating the low price $l$. 

If this simple market system stays in the mixed-strategy Nash equilibrium, the
mixed strategy $(w_l^0, w_m^0, w_h^0)$ for each shop to choose the low, middle,
and high price is
\begin{equation}
  w_l^0  =  \frac{m h - 3 l (h-m)}{m l + (m-l) h} \; , \quad
  w_m^0  =  \frac{ (m+l) h - 3 m l}{m l + (m-l) h} \; , \quad
  w_h^0  =  \frac{m l - (m-l) h}{m l + (m - l) h} \; .
\end{equation}
Notice that $w_l^0 > w_m^0 > w_h^0$, meaning that each shop should choose the
low price with the highest probability and the high price with the lowest
probability. For the representative parameter set $l=1$, $m= 1.45$ and $h=2.1$,
we have $\rho_l^0 \approx 0.457$, $\rho_m^0 \approx 0.332$ and
$\rho_h^0 \approx 0.211$. The expected payoff per day (normalized by $N$) is
$g^0 = m h l/[m l + (m - l) h]$, which is always larger than $(6/5) l$ but less
than $(4/3)l$. 

But in reality the prices of different shops are highly entangled
\cite{Lach-2002,Noel-2007}. For our two-shop toy model, a simple microscopic
process to mimic correlated decision-making is the noisy best-response dynamics
with a single parameter $\beta$ \cite{Blume-1993}. Knowing the price
$s_{t-1}^{(x)}$ of shop X at day $(t-1)$ but uncertain of its price at the next
day $t$, shop Y chooses its price $s_t^{(y)} \in \{l, m, h\}$ at day $t$
according to the conditional probability distribution
\begin{equation}
  \label{eq:Blume}
  P(s_t^{(y)} | s_{t-1}^{(x)}) = \frac{e^{\beta g(s_t^{(y)} | s_{t-1}^{(x)})}}{
    e^{\beta g( l | s_{t-1}^{(x)})} + e^{\beta g(m | s_{t-1}^{(x)})}
    + e^{\beta g(h | s_{t-1}^{(x)})}} \; ,
\end{equation}
where  $g(s^{(y)} | s^{(x)})$ is shop Y's payoff (normalized by $N$) in state
$s^{(y)}$ while the other shop X chooses price $s^{(x)}$, e.g., $g(l | l) = l$,
$g(m | l) = m/2$, and $g(h | l) = h/2$.

\begin{figure}[t]
  \centering
  \includegraphics[angle=270,width=1.0\textwidth]{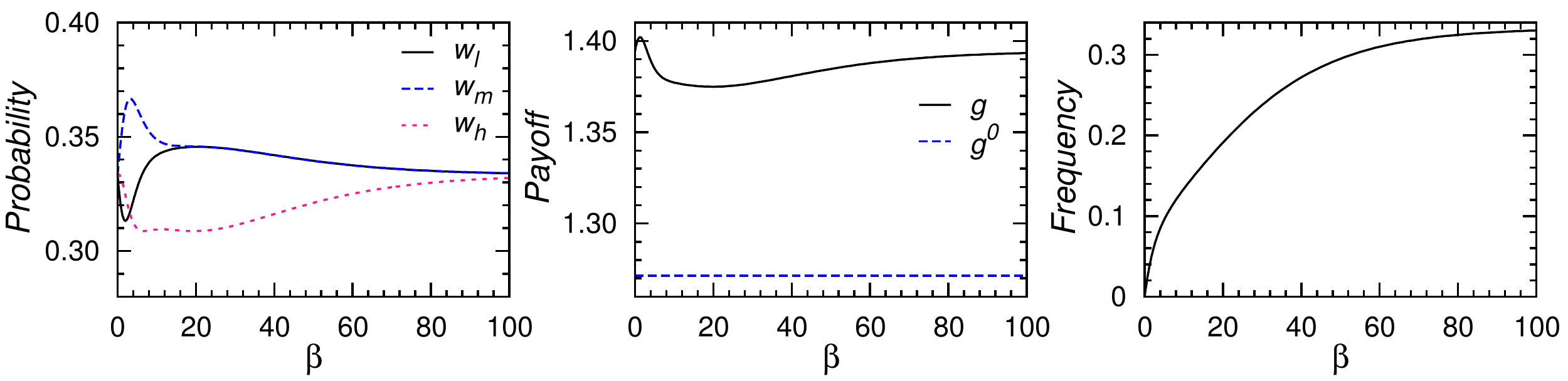}
  \caption{ \label{fig:price}
    The noisy best-response model of price competition with response parameter
    $\beta$. The two shops can choose among the low price $l=1$, the medium
    price $m=1.45$ and the high price $h=2.1$. (left) The steady-state
    probabilities $w_l$, $w_m$ and $w_h$ of choosing the three prices, which
    are quite different from the corresponding values
    ($w_l^0 \approx 0.457$, $w_m^0 \approx 0.332$, $w_h^0 \approx 0.211$) of the
    mixed-strategy Nash equilibrium. (middle) The expected payoff per game round
    $g$ as compared to the NE value $g^0$. The maximal value $g \approx 1.402$
    is reached at $\beta \approx 1.3$. (right) The mean cycling frequency of the
    system's social state. }
\end{figure}

The steady-state marginal probabilities ($w_l$, $w_m$, and $w_h$) of choosing
the three prices are easy to compute if both shops are governed by the same
stochastic dynamics (\ref{eq:Blume}). This steady-state distribution is very
different from the NE mixed strategy (see Figure~\ref{fig:price}, left). If
$\beta=0$, there is no selection, $w_l=w_m=w_h=1/3$, and the expected payoff
per day is $g = (2 h + 3 m + 4 l)/9$. As $\beta$ increases the expected payoff
changes in a non-monotonic way, and there is an optimal value of $\beta$ at
which the expected payoff reaches
the global maximum value much higher than the NE value
$g^0$  (Figure~\ref{fig:price}, middle).

We can study the collective behaviour of this two-shop toy model following the
same method of Section~\ref{subsec:RPSobserv6}. When $\beta >0$ there is
persistent cycling in the system (Figure~\ref{fig:price}, right), which is consistent
with the Edgeworth price cycle \cite{Hopkins-Seymour-2002,Edgeworth-1925}. The
cycling frequency is an increasing function of $\beta$ (positive cycling
direction is high price $\rightarrow$ middle price $\rightarrow$ low price
$\rightarrow$ high price). 

Real-world markets are of course much more complex than the model discussed
here. The key points we want to emphasize are (i) cyclic dominance among
different price levels do occur in real markets \cite{Lach-2002,Noel-2007},
and (ii) it is much more beneficial to make intuition-guided decisions rather
than to follow the Nash-equilibrium mixed strategy in price competition. The
second point is indeed closely related to the debated issue of rationality (the
reader may consult \cite{Basu-2007} for more discussions).

\section{Outlook}
\label{sec:outlook}

The Rock-Paper-Scissors game is a simple game that helps improving our
understanding on many complex competition issues (species divergence, price
cycling, human decision-making, rationality and cooperation and so on). This
game is the simplest model system for studying the non-equilibrium statistical
mechanics of non-cooperative strategic interactions, and it can serve as a
starting point to enter into the interdisciplinary field between statistical
physics and game theory.

In this brief review we have left untouched the issue of
possible phase transitions. If the Rock-Paper-Scissors game is played on an
infinite lattice or a complex network, are there competition-driven critical
phenomena and how to quantitatively describe these behaviours?
At present, how players adjust their decision-making parameters is still quite
unclear and is largely ignored in theoretical investigations. A lot of
empirical and theoretical efforts are needed in these directions.

\section*{Acknowledgement}
I thank Prof. Bin Xu and Dr. Zhijian Wang for collaborations, and thank
Prof. Bin Xu for a critical reading of the manuscript. This work is partially
supported by the National Basic Research Program of China (grant number 
2013CB932804) and by the National Natural Science Foundations of China (grant
numbers 11121403 and 11225526).

\vskip 0.6cm
Hai-Jun Zhou received his Doctorate from the Chinese Academy of Sciences in 
2000. His dissertation was on structural transitions in biopolymers.
Since 2005 he has been a research professor at the Institute of Theoretical
Physics of the Chinese Academy of Sciences. Currently he is working on
statistical physics and its interdisciplinary applications, including spin
glasses and combinatorial optimization, percolation processes in complex
networks, and collective dynamics in game systems.


\end{document}